\documentclass[11pt]{asaproc}

\usepackage{algorithm}
\usepackage[noend]{algorithmic}
\usepackage{amsfonts}
\usepackage{amsmath}
\usepackage{amsrefs}
\usepackage{amssymb}
\usepackage{array}
\usepackage{bm}
\usepackage{caption}
\usepackage{chngpage}
\usepackage{cite}
\usepackage{color}
\usepackage{enumerate}
\usepackage{epsf}
\usepackage{epsfig}
\usepackage{framed}
\usepackage{fullpage}
\usepackage{graphicx}
\usepackage{latexsym}
\usepackage{lineno}
\usepackage{lipsum}
\usepackage{multicol}
\usepackage{setspace} 
\usepackage{stfloats}
\usepackage{times}
\usepackage{txfonts}
\usepackage{ulem}
\usepackage{url}
\usepackage{verbatim}
\usepackage[table]{xcolor}
\usepackage{xr}

\algsetup{indent=2em}

\linenumbers



\title{A Hierarchical Bayesian  Approach 
\\for  Aerosol Retrieval Using MISR Data}
\author{Yueqing Wang $^{1,\footnote{Corresponding author: yqwang@stat.berkeley.edu.}}$, Xin Jiang $^{2,\footnote{Xin Jiang is now working at Netease Youdao.  1 Zhongguancun East Road, Haidian District, Beijing, 100084 China.}}$ , Bin Yu $^{1, 3}$, Ming Jiang $^{2, 4}$
}
\date{July 2011}

\begin{document}

\pagestyle{empty}
\maketitle
\begin{center}
$^1$  Department of Statistics, University of California at Berkeley, CA 94720-3860, U.S.

$^2$  LMAM, School of Mathematical Sciences, Peking University, Beijing 100871, China.

$^3$  Department of Electrical Engineering and Computer Sciences,\\ University of California at Berkeley, CA 94720-3860, U.S.

$^4$  Beijing International Center for Mathematical Research, Beijing, 100871, China.
\end{center}

\begin{abstract}
Atmospheric aerosols can cause serious damage to human health and life expectancy.  Using the radiances observed by NASA's Multi-angle Imaging SpectroRadiometer (MISR), the current MISR operational algorithm retrieves Aerosol Optical Depth (AOD) at 17.6 km resolution.  A systematic study of aerosols and their impact on public health, especially in highly-populated urban areas, requires finer-resolution estimates of AOD's spatial distribution.

We embed MISR's operational weighted least squares criterion and its forward calculations for AOD retrievals in a likelihood framework and further expand into a hierarchical Bayesian model to adapt to finer spatial resolution of 4.4 km. To take advantage of AOD's spatial smoothness, our method borrows strength from data at neighboring areas by postulating a Gaussian Markov Random Field prior for AOD.  Our model considers AOD and aerosol mixing vectors as continuous variables, whose inference is carried out using Metropolis-within-Gibbs sampling methods.  Retrieval uncertainties are quantified by posterior variabilities.  We also develop a parallel MCMC algorithm to improve computational efficiency.  We assess our retrieval performance using ground-based measurements from the AErosol RObotic NETwork (AERONET) and satellite images from Google Earth.

Based on case studies in the greater Beijing area, China, we show that   4.4 km resolution can improve both the accuracy and coverage of remotely-sensed aerosol retrievals, as well as our understanding of the spatial and seasonal behaviors of aerosols.  This is particularly important during high-AOD events, which often indicate severe air pollution.

\end{abstract}

\section{Motivation}\label{sec:moti}

Atmospheric aerosols, complex mixtures of solid particles and liquid droplets in the air, can significantly affect human health and life expectancy \cite{poschl2005}.  
When inhaled, aerosols can penetrate cell membranes, then migrate and seriously damage human respiratory, cardiovascular systems \cite{pope2002} and the brain \cite{monleau2005}.
Short-term impacts include: irritation to eyes, nose and throat; upper respiratory infections including pneumonia and bronchitis; and stroke or death from cardiovascular causes. 
Continual exposure to hazardous aerosols can aggravate or complicate medical conditions in the elderly\cite{degouw2011}; aerosols from silica and diesel can lead to diseases including silicosis and black lung.   
Aerosols with an aerodynamic diameter less than 2.5~$\mu$m, such as black carbon, can severely reduce ground-level visibility.  
Profiling spatial distribution of aerosols at fine resolution is thus critical for air quality and public health studies, especially in urban areas with complex anthropogenic aerosol sources, such as vehicles, power plants, and factories that burn fossil fuels.  

There are two approaches to measure the spatial distribution of aerosols: through ground-based measurements or remote-sensed radiance imageries.
Both quantify the amount of aerosols by spectral Aerosol Optical Depth (AOD), defined as the negative logarithm of the fraction of radiation (sunlight) not scattered or absorbed by aerosols on a path in the Earth's atmosphere\footnote{For example, an AOD value of 2.5 corresponds to 92$\%$ of radiation scattered or absorbed.}.
AOD at different spectral bands can be viewed as known functions of AOD at the green band using the Angstr$\ddot{o}$m power law\cite{liou2002}.  		 
	For notational simplicity, this paper refers to AOD at the green band.
With either ground or remote-sensing approach, the spatial and temporal variabilities of aerosols require continual observations and computationally efficient analyses.

\begin{figure}[htp]
\centering
\resizebox{0.315\columnwidth}{!}{\includegraphics{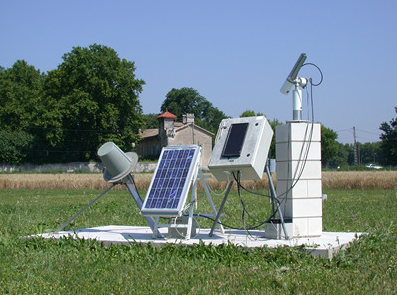}}
\resizebox{0.3\columnwidth}{!}{\includegraphics{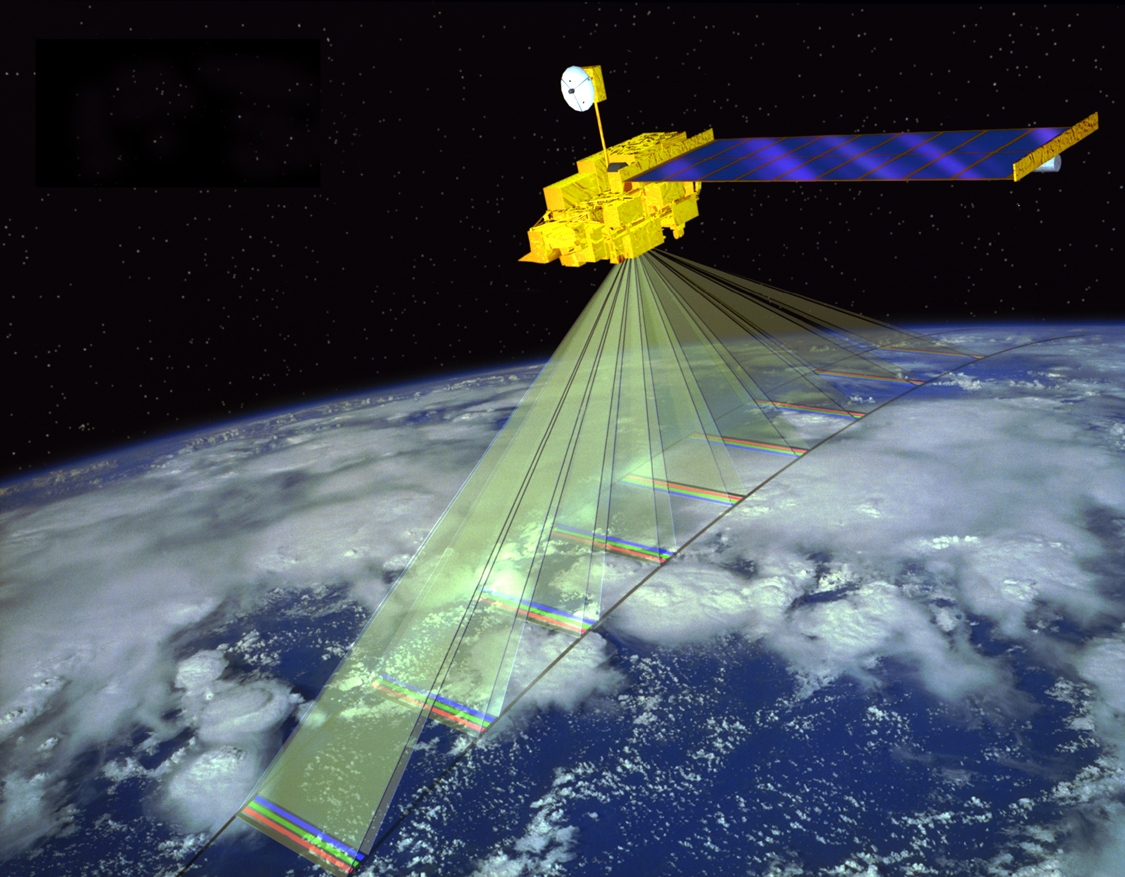}}
\caption{\label{fig:camera}AERONET sun photometer at Avignon, France (left) and MISR cameras (right).}
\end{figure}

The AErosol RObotic NETwork (AERONET)\cite{AERONET:online} provides a data archive of local AOD values using a network of automatic sun photometers (Figure \ref{fig:camera}, left panel) 
located at more than 400 stations on the Earth's surface.  
It measures AOD from every half hour to every two hours, with uncertainties $<\pm 0.01$ at wavelengths $> 440$ nm\cite{holben1998}.  
AERONET measurements are  widely accepted as a gold standard to validate AOD estimates based on other data sources.  
The sparse and heterogeneous locations of AERONET stations, however, make it difficult to  directly use their measurements to study the spatial behaviors of aerosols.

Remote-sensing radiometers offer a better spatial coverage by retrieving AOD from radiance imageries over the Earth's entire surface, such as the Multi-angle Imaging SpectroRadiometer (MISR) aboard the NASA Earth Observing System Terra satellite (Figure \ref{fig:camera}, right panel).  
MISR views the day-lit Earth atmosphere almost simultaneously at nine angles along its track.  
This unique design of multiple viewing angles provides an enhanced sensitivity to aerosol scattering and cloud reflective effects\cite{diner1998}, rendering MISR a significant advantage over other remote-sensing instruments.  
MISR outputs four-spectral imageries at 1.1~km resolution for the blue, green, near-infrared bands, and at 275 m for the red band.  Based on these imageries, MISR then produces AOD retrievals at 17.6 km resolution.  
To quantitatively represent aerosol mixtures, aerosol particles are  characterized and categorized according to their properties such as radius and single scattering albedo (SSA)\footnote{SSA is defined as the ratio of scattered radiation to total extinct radiation (scattered and absorbed).}.  
Each category is referred to as a component aerosol.
Then an aerosol mixture is identified by a notion of composition: a collection of $M$ component aerosols and their mixing vector relative to these $M$ components.  
Elements of the $M$-dimensional mixing vector sum up to 1, indicating  mixing percentages of the $M$ components.
To simplify  remote-sensing retrieval, MISR operational algorithm considers only 21 component aerosols and 74 pre-fixed compositions\footnote{The number of non-zero elements of the MISR's 74 mixing vectors are no more than three.}.
	Based on the known physical and compositional properties of each component aerosol, forward radiative transfer calculations are performed to provide atmospheric radiation field in the 36 MISR channels (9 viewing angles $\times$ 4 spectral bands).
The results are stored in the Simulated MISR Ancillary Radiative Transfer (SMART) Dataset.  
The MISR operational aerosol retrieval algorithm adopts a weighted least squares criterion to determine whether the radiative transfer calculated radiances provide good fits to the MISR-observed radiances.  
Validated by AERONET measurements,  MISR field measurements, and airplane campaigns\cite{diner2002}, MISR's retrievals have shown to be informative in characterizing aerosols' optical properties.  Previous studies include those on wildfire smoke\cite{kahn2008}, mineral dusts\cite{koven2008}, and climate changing aerosols\cite{solmon2006}.  

MISR's ability to capture aerosol-related information makes it well suited  to assist studies on aerosols' impact on public health.
However, the heterogeneity of urban aerosols within an area of 17.6 $\times$ 17.6 km$^2$, the spatial resolution of MISR AOD retrievals, makes finer resolution desirable.  For example, San Francisco is represented by less than half of a MISR pixel.
Yet the residents of San Francisco are  exposed to varying levels of air pollution.
Case studies in Delhi show that 5-km AOD has a significantly higher association with health-related particulate matters than AOD of rougher resolution \cite{kumar2007}.
As a result, we use 4.4 km as our retrieval resolution, also to be compatible with the MISR observations at 1.1 km.
Also, observational studies indicate that tropospheric aerosol burden has increased at mid-latitudes and in the Arctic, probably due to anthropogenic activities\cite{peterson1981}\cite{shaw1982}.
This suggests that more varieties beyond the 74 pre-fixed aerosol compositions are to be considered in order to capture aerosols' growing heterogeneity.  

	Finer-resolution retrievals with greater varieties of aerosol compositions lead to a larger number of parameters to estimate.  This is possible if we take advantage of AOD's spatial smoothness and reduce the 21 component aerosols to a smaller subset, say four, chosen according to current knowledge of the study region's aerosol conditions.  
In particular, a hierarchical Bayesian  model is proposed to retrieve AOD values and mixing vectors based on MISR observations at 4.4 km resolution.
We adopt a likelihood framework based on MISR's weighted least squares and construct the Bayesian hierarchy to incorporate AOD's spatial smoothness using a Gaussian Markov Random Field (GMRF) prior.
The movement and dispersion of air particles in the  atmosphere justify the spatial smoothness of AOD from a physical viewpoint.
	To flexibly describe various aerosol conditions, our model regards AOD  values and mixing vectors as continuous parameters.  This expands the set of possible  compositions beyond the 74  pre-fixed choices of MISR.  We show how this enriched variety is necessary to retrieve heterogeneous urban aerosols.
Our study takes a MISR Block\footnote{MISR observes the Earth's surface in 233 swaths; each swath contains 180 560 $\times$ 140 km$^2$ MISR Blocks.} as a data unit to balance the coverage of a greater metropolitan area and computational cost.

The posterior inference of AOD and mixing vectors is carried out using Markov Chain Monte Carlo (MCMC) sampling methods, particularly Metropolis-within-Gibbs.  Such sampling methods allow us to quantify the retrieval uncertainties by posterior variabilities.  The algorithm, however, is computationally intense.  
	We develop a parallel MCMC algorithm by partitioning a MISR Block into smaller patches, in order to enable parallel samplings while maintaining the overall smoothness level using summary statistics.  We show that retrievals from the two algorithms are consistent, with an increase in computational speed for the parallel MCMC algorithm.
	To assess the performance of our methods, we apply them to retrieve AOD values for the greater Beijing area in China.  Our retrievals are tested against ground-based measurements of AOD from two AERONET stations in the area.  Results show improvement on retrieval accuracy and coverage, especially during high-AOD events.  We also include geographical conditions and levels of anthropogenic activities from Google Earth to qualitatively validate our results.  

The rest of the paper is organized as follows:  Section \ref{sec:model} provides the rationale and details of our Bayesian model for retrieving AOD values and mixing vectors, while Section \ref{sec:alg}  details our MCMC algorithms.  
Section \ref{sec:res} contains  case studies for model validation and  interpretation, comparing our results with MISR's retrievals and AERONET measurements.
Section \ref{sec:comp} illustrates the necessity to include a richer variety of aerosol compositions.  
Section \ref{sec:dis} summarizes the results and suggests directions for future research.

\section{Hierarchical Bayesian Model}\label{sec:model}

Our objective is to establish a more detailed data-driven description of the relationship among radiances, AOD, and aerosol compositions to assist aerosol-related health studies.  The MISR operational retrieval algorithm provides this information by comparing the observed and the radiative transfer calculated radiances, but it is limited within the 74 pre-fixed aerosol compositions and a discrete grid of AOD values.
We propose to allow a greater variety of aerosol optical behaviors by considering AOD values and mixing vectors as continuous variables, given a fixed set of four component aerosols.
For the greater Beijing area, this set includes spherical non-absorbing aerosols without sulfate, spherical non-absorbing aerosols with sulfate, spherical absorbing aerosols, and grains (dust).

Each MISR Block contains 256 pixels (8 rows $\times$ 32 columns) at 17.6 km resolution in the MISR retrievals.  The number of pixels in a MISR Block rises to 4,096 (32 rows $\times$ 128 columns) at 4.4 km resolution, presenting a more complex problem with approximately 16,384 parameters to estimate. 	 
	On the other hand, air particles interact in the  atmosphere within a certain range; they affect aerosol conditions in near neighborhoods\cite{bergametti1989}\cite{winchester1971}.  
	This suggests a stronger spatial dependence among adjacent pixels at a finer scale.
	When modeling at fine resolution, therefore, it is  necessary and beneficial to borrow strength from AOD's spatial smoothness to reduce model complexity. 
In particular, we construct a hierarchical Bayesian  model with a built-in spatial dependence using a Gaussian Markov Random Field prior for AOD.

\subsection{Defining the Likelihood Function}\label{sec:lik}

	Let $p=1,\dots,P$ index the $P=4,096$ pixels on a two-dimensional lattice in a MISR Block at 4.4 km resolution, and $\bm{L} = (\bm{L}_1, \dots, \bm{L}_P)$ denote the MISR-observed top-of-atmosphere radiances.
	For each pixel~$p$, $\bm{L}_p=(L_{1p},\dots,L_{Cp})\in\mathbb{R}^C$ corresponds to MISR's $C=36$ channels.
	For every channel $c=1,\dots,C$, the MISR retrieval algorithm sets a measurement error of size $\sigma_c$ as $5\%$ of the smaller value between 0.04 and $\bar{L}_c = (\sum_{p=1}^P L_{cp})/P$.
	For pixel $p$, our goal is to estimate its AOD value $\tau_p\in\mathbb{R}$ and mixing vector $\bm{\theta}_p = (\theta_{p1},\dots, \theta_{pM})\in\mathbb{R}^M$, relative to the $M$ component aerosols involved ($ \bm{\theta}_p\geq 0$ and $\sum_{m=1}^M\theta_{pm} = 1$).
	Each of MISR's 74 pre-fixed aerosol mixtures contain two or three component aerosols.  We expand to allow mixtures of four component aerosols by setting $M=4$; case studies confirm the sufficiency of this choice.
	
	Given the geolocation of pixel $p$, its AOD value $\tau_p$, a set of component aerosols and their mixing vector~$\bm{\theta}_p$, Radiative Transfer (RT) equations are used to simulate radiances  $\bm{L}^{RT}=(L_1^{RT},\dots,L_C^{RT})$ \cite{diner1999}; their pre-computed values at discrete points are stored in MISR's SMART Dataset\footnote{The other parameters, such as the ambient pressure, take the default values unless otherwise specified.  The MISR team has kindly given us access to the SMART dataset.}.
	Thus, $\bm{L}^{RT}$ can be viewed as functions of $(\tau_p,\bm{\theta}_p)$, relative to the $M$ component aerosols involved.  
	For each pixel $p$ independently, the MISR operational retrieval algorithm uses a weighted least squares criterion to measure the closeness of an observed radiance vector to a particular RT simulated radiance vector.  The weighted least squares take the following 
	form \cite{diner2008}:
	\begin{equation}\label{eq:chisq2}
	\chi_p^2 = \sum_{c=1}^{C}\frac{(L_{cp}-L_c^{RT}(\tau_p,\bm{\theta}_p))^2}{2\sigma_c^2}.
	\end{equation}

	The MISR retrieval algorithm exhaustively searches over all combinations of pre-fixed AOD values and 74 aerosol compositions to match $\bm{L}^{RT}$ to the observed $\bm{L}$. 
	The combinations of AOD and compositions satisfying a pre-established threshold of $\chi_p^2$ in \eqref{eq:chisq2} are considered good fits to the observations; the average of all such AODs  is the MISR retrieval at pixel $p$.
	
	Inspired by MISR's weighted least squares criterion, we propose to use the weighted differences between observed $\bm{L}$ and radiative transfer simulated $\bm{L}^{RT}$ in \eqref{eq:chisq2} to form the following operational likelihood function:
\begin{equation}\label{eq:likelihood}
p(\bm{L}|\bm{\tau},\bm{\theta})\propto \exp\left\{-\sum_{c=1}^{C}\sum_{p=1}^P\frac{(L_{cp}-L_c^{RT}(\tau_p,\bm{\theta}_p))^2}{2\sigma_c^2}\right\}.
\end{equation}
	 If we carry out a Maximum Likelihood estimation, the above Gaussian likelihood function coincides with MISR's weighted least squares criterion, assessing how relatively probable are the unobserved parameters $\bm{\tau}=(\tau_1, \dots, \tau_P)$ and $\bm{\theta}=(\bm{\theta}_1, \dots, \bm{\theta}_P)$, given the MISR observations $\bm{L}$.  
	More importantly, this operational likelihood provides a formal device for us to construct a spatial smoothness structure for the AOD values $\bm{\tau}$ into the Bayesian hierarchy.
	
	Even though the exact distribution of the weighted differences in \eqref{eq:likelihood} is difficult to determine due to the complex origins for these differences\footnote{Such origins include MISR camera measurement errors, radiative transfer calculation noises, differences between the proposed and true values for AOD and mixing vectors, choices of component aerosols, and errors in estimating  surface-leaving radiances.}, histograms of retrieval residuals based on \eqref{eq:likelihood} display a single modal distribution; this supports our choice for a Gaussian-shaped operational likelihood.
	Another assumption in both \eqref{eq:chisq2} and \eqref{eq:likelihood} is that the differences between $\bm{L}^{RT}$ and $\bm{L}$ are independent of the channel $c$\footnote{We found close-to-0 correlations (-0.0445) between our retrievals' residuals at different viewing angles, but nontrivial correlations (0.5714) between residuals at different spectral bands.  In current work, we are building this dependence structure among different bands in our model.}, if the correct values of ($\bm{\tau}$, $\bm{\theta}$) have been selected.

	Now we are ready to describe our hierarchical model through building conditional relationships within the Bayesian hierarchy and assigning reasonable priors to the unobserved variables.

\subsection{\label{cdetails}Construction of Priors and Conditional Probabilities}

For fixed atmospheric pressures, humidity, wind levels, and a set of component aerosols involved, the top-of-atmosphere radiances $\bm{L}$ are mainly determined by AOD $\bm{\tau}$ and aerosol mixing vectors $\bm{\theta}$.  Our Bayesian hierarchy's first level depicts this dependence of $\bm{L}$ on $\bm{\tau}$ and $\bm{\theta}$.  
	Prior distribution for $\bm{\tau}$ is postulated to capture the spatial smoothness, calibrated by hyperparameter $\kappa$. We further assume independence between priors for $\bm{\tau}$ and $\bm{\theta}$ to simplify computation, i.e., $p(\bm{\tau}, \bm{\theta})=p(\bm{\tau})p(\bm{\theta})$.  
The inference of parameters and hyperparameters using MCMC sampling methods, is discussed in Section \ref{sec:alg}.

\subsubsection{\textbf{Prior Beliefs about AOD's Spatial Dependence}}\label{sec:kapp}

We characterize the spatial dependence of AOD values $\bm{\tau}$ using an intrinsic Gaussian Markov Random Field (GMRF) prior of first order\cite{rue2005}.  Define $\kappa$ as the homogenous scaler precision and use $\sim$ to indicate spatial adjacency. The following prior is invariant to perturbation by the same constant to $\bm{\tau}$ of all pixels\cite{besag1974},
\begin{equation}\label{eq:tauprior}
p(\bm{\tau}|\kappa) \propto\kappa^{\frac{P-1}{2}}\exp\left\{-\frac{\kappa}{2}\sum_{p^{\prime}\sim p}(\tau_{p^{\prime}}-\tau_p)^2\right\}.
\end{equation}
This allows us to model AOD's spatial smoothness by penalizing sharp changes of $\bm{\tau}$ among adjacent pixels, regardless of their unknown overall level.
The prior in \eqref{eq:tauprior} is calibrated by $\kappa$ as AOD's precision.   The larger $\kappa$ is, the smoother the region's AOD values are.
For some regions, however, a more complicated GMRF prior might be necessary.  For example, a constant wind pattern might require distinguishing an upwind pixel from a downwind pixel. This paper works with a homogenous precision $\kappa$ and thus has its limitations.

To estimate $\kappa$, we assign it a hyperprior.
	Due to AOD's large variability within a day and the lack of pre-existing records to specify a prior belief of $\bm{\tau}$'s behaviors, we consider a noninformative prior: $p(\kappa)\propto 1/\kappa$.  The posterior is a proper Gamma distribution,
\begin{equation}
p(\kappa|\bm{\tau})\propto \kappa^{\frac{P-1}{2}-1}\exp\left\{-\frac{\kappa}{2}\sum_{p^{\prime}:p^{\prime}\sim p}(\tau_{p^{\prime}}-\tau_p)^2\right\}.
\end{equation}
For the above prior to work well, the number of groups, namely $P$, is to be larger than 5 \cite{gelman2006}. In our case, $P$ is commonly larger than 1000 at 4.4 km resolution.  
The simulation to be described in Section \ref{sec:sampler} shows good agreement between the true and retrieved values of $\kappa$ using our MCMC algorithm. For example, we observed 100 (true) and 92.08 (retrieved) in one simulation, 500 (true) and 485.76 (retrieved) in another.

\subsubsection{\textbf{Prior Specification for Aerosol Compositions}}\label{sec:thetap}

Prior information on aerosol compositions is incorporated in the model through choices of the $M=4$ component aerosols involved, based on  geophysical knowledge of the study region. To model the mixing vectors $\bm{\theta}$ of the $M$ component aerosols, we use an $M$-dimensional Dirichlet prior with Dirichlet parameter $\bm{\alpha}=(\alpha_1, \dots, \alpha_M)$. 
Conditioning on $\bm{\alpha}$, the mixing vectors $\{\bm{\theta}_p\}_{p=1}^P$ are considered to be independent of each other,
\begin{equation}\label{eq:thetaprior}
p(\bm{\theta}|\bm{\alpha}) = \prod_{p=1}^Pp(\bm{\theta}_p|\bm{\alpha}) = \prod_{p=1}^P\frac{\Gamma(\sum_{m=1}^M\alpha_m)}{\prod_{m=1}^M\Gamma(\alpha_m)}\theta_{p1}^{\alpha_1-1}\cdots\theta_{pM}^{\alpha_M-1}.
\end{equation}
Even though the mixing vectors' spatial smoothness is not explicitly formulated, it is still captured and implicitly enforced by the spatial structure of AOD $\bm{\tau}$ through their dependence on the observed radiances~$\bm{L}$.  In fact, our estimates of mixing vectors $\bm{\theta}$ indeed display spatial smoothness.
	The model and algorithms remain relatively simple and computationally efficient. 

We can further control the overall sparsity of the mixings of component aerosols by adjusting the magnitude of $\bm{\alpha}$.  In general, we obtain no prior information on the mixing's sparsity; we assign $\bm{\alpha}$ a hyperprior to estimate it.  
Since \eqref{eq:thetaprior}  belongs to an exponential family, we adopt its conjugate: $p(\bm{\alpha})\propto \exp(\sum_{m=1}^M(1-\alpha_m))$. 
	This prior of $\bm{\alpha}$ gives larger probability to a smaller sum of $\alpha_m$'s, which suggests a sparse mixing of component aerosols, i.e. mixtures with one or two dominant components.  This is supported by results from observational studies on aerosol mixings\cite{diner1999}.  The posterior has the following form,

\begin{equation*}
	p(\bm{\alpha}|\bm{\theta})\propto\exp\left\{\sum_{m=1}^M(\alpha_m-1)(\sum_{p=1}^P\log\theta_{pm}+1) -P\times(\sum_{m=1}^M\log\Gamma(\alpha_m)-\log\Gamma(\sum_{m=1}^M\alpha_m))\right\}.
\end{equation*}

\subsubsection{\textbf{Hyperprior for $\bm{\sigma}^2$}}\label{sec:hysig}

	In our approach,  we regard $\{\sigma_c^2\}_{c=1}^{C}$ as unknown and they are estimated together with ($\bm{\tau}$, $\bm{\theta}$).
	The likelihood function for $\bm{\sigma}^2=(\sigma_1^2,\dots,\sigma_C^2)$, $p(\bm{L}|\bm{\sigma}^2,\bm{\tau},\bm{\theta})$, follows a normal distribution with known mean and unknown variance.  
We adopt a noninformative scaled inverse-$\chi^2$ hyperprior for $\bm{\sigma}^2$ to model the channel weights $\{\frac{1}{2\sigma_c^2}\}_{c=1}^{C}$: $p(\sigma_c^2)\propto\sigma_{c}^{-2}$. 
	This hyperprior suggests that values for the unknown weights become less likely in inverse proportion to their values; it is also a choice of computational convenience.  The conditional posterior also follows the scaled inverse-$\chi^2$ distribution,

\begin{equation*}
p(\sigma_c^2|\bm{\tau},\bm{\theta},\bm{L})\propto (\sigma_c^2)^{-(\frac{P}{2}+1)}\exp\left\{-\frac{\sum_{p=1}^P(L_{cp}-L_c^{RT}(\tau_p,\bm{\theta_p}))^2}{2\sigma_c^2}\right\}.
\end{equation*}

\section{MCMC Retrieval Algorithms}\label{sec:alg}
 
Based on the hierarchical Bayesian model  previously developed, this section first derives marginal posterior distributions of AOD values $\bm{\tau}$ and mixing vectors $\bm{\theta}$.  We then devise two MCMC algorithms to sample from the posteriors. Using MISR observed radiances as input, we take the sampled posterior means as outputs.

\subsection{Posterior Distributions of AOD Values and Mixing Vectors}\label{sec:postd}

The full Bayesian model discussed above can be summarized as follows:
\begin{align*}
	\bm{L}_p | \tau_p, \bm{\theta_p} &\sim \mathcal{N}(\bm{L}^{RT}(\tau_p,\bm{\theta}_p), \bm{\sigma^2}), \mbox{  } p = 1, \dots, P,\\
	\bm{\tau} | \kappa &\sim GMRF(\kappa),\\
	\bm{\theta} | \bm{\alpha} &\sim Dirichlet(\bm{\alpha}),\\
			\bm{\sigma}^2 &\sim scaled \mbox{ } inverse-\chi^2 (\nu_0),\\
	\kappa &\sim Gamma(\alpha_0, \beta_0),\\
	p(\bm{\alpha}) &\sim Exp(\sum_{m=1}^M(1-\alpha_m)).
\end{align*}
With no additional information on the hyperparameters, $\nu_0$, $\alpha_0$, and $\beta_0$ are chosen to be 0 for convenience and later shown to be robust.
The marginal posterior  of AOD values $\bm{\tau}$ is,
\begin{equation}\label{eq:taucon}
p(\bm{\tau}|\bm{\theta}, \kappa, \bm{\sigma}^2, \bm{L}) \propto 
 \exp\left\{-\frac{1}{2}\kappa\sum_{p^{\prime}:p^{\prime}\sim p}(\tau_{p^{\prime}}-\tau_p)^2-\sum_{c=1}^C\sum_{p=1}^P\frac{(L_{cp}-L^{RT}_c(\tau_p,\bm{\theta}_p))^2}{2\sigma_c^2}\right\}.
\end{equation}
The marginal posterior distribution of the mixing vectors $\bm{\theta}$ can be expressed as,
\begin{equation}\label{eq:thetacon}
p(\bm{\theta}|\bm{\tau}, \bm{\alpha}, \bm{\sigma}^2, \bm{L})\propto\exp\left\{\sum_{p=1}^P\sum_{m=1}^M(\alpha_m-1)\log\theta_{pm}-\sum_{c=1}^C\sum_{p=1}^P\frac{(L_{cp}-L^{RT}_c(\tau_p,\bm{\theta}_p))^2}{2\sigma_c^2}\right\}.
\end{equation}
Both posteriors contain radiative transfer simulated $\bm{L}^{RT}$, which can be obtained at necessary values through interpolations from the MISR SMART Dataset, using $\tau_p$ and $\bm{\theta}_p$ as inputs \cite{diner1999}.  The resulted non-closed-form posteriors, however, are difficult to directly sample from.
A Metropolis-within-Gibbs sampler is thus used.

\subsection{Metropolis-within-Gibbs Sampling from the Posterior Distributions}\label{sec:sampler}

The Gibbs sampler\cite{geman1984} is a numerical technique to sample from a joint distribution, $p(\bm{\tau}, \bm{\theta}, \bm{\sigma}^2,\kappa,\bm{\alpha}|\bm{L})$ in our case.
We sample for $\tau_p$ and $\bm{\theta}_p$ using a Metropolis-Hastings (M-H) sampler, for each pixel $p$ on the MISR Block column by column and pixel by pixel.  The following proposal distribution is used in M-H sampler for $\tau_p$,

\begin{equation*}
	p(\tau_p|\bm{\tau}_{-p}) \propto \exp\left(-\frac{n_p\kappa}{2}(\tau_p-\frac{1}{n_p}\sum_{p^{\prime}:p^{\prime}\sim p}\tau_{p^{\prime}})^2\right),
\end{equation*}
where $n_p$ is the number of adjacent pixels to pixel $p$, and $\kappa$ the scalar precision of the Markov Random Field. 
A Dirichlet proposal distribution with parameter $\bm{\alpha}$ is used in M-H for $\bm{\theta}_p$.
Denote vector $(\tau_p,\dots, \tau_{p^{\prime}})$ by $\bm{\tau}_{p:p^{\prime}}$ and similarly for $\bm{\theta}$ and their Dirichlet parameter $\bm{\alpha}$.
	Given initializations $(\bm{\tau}^{(0)}, \bm{\theta}^{(0)}, (\bm{\sigma}^2)^{(0)}, \kappa^{(0)},\bm{\alpha}^{(0)})$, the sampler proceeds as described in the following Metropolis-within-Gibbs Algorithm.

\begin{algorithm}
	\caption*{\textbf{Metropolis-within-Gibbs Algorithm (M-w-G)}}
	At step $t$, iterate the following process:
	\begin{algorithmic}[1]
		\FOR {$p=1$ to $P$}
		\STATE Use M-H to sample $\tau_p^{(t)} \sim p(\tau_p | \bm{\tau}_{1:(p-1)}^{(t)}, \bm{\tau}_{(p+1):P}^{(t-1)}, \bm{\theta}^{(t-1)}, (\bm{\sigma}^2)^{(t-1)},\kappa^{(t-1)}, \bm{L})$.
		\ENDFOR
		\FOR {$p=1$ to $P$}
		\STATE Use M-H to sample $\bm{\theta}_p^{(t)} \sim p(\bm{\theta}_p | \bm{\tau}^{(t)}, \bm{\theta}_{1:(p-1)}^{(t)}, \bm{\theta}_{(p+1):P}^{(t-1)}, (\bm{\sigma}^2)^{(t-1)},\bm{\alpha}^{(t-1)}, \bm{L})$.
		\ENDFOR
		\FOR {$c=1$ to $C$}
		\STATE Use M-H to sample $(\sigma_c^2)^{(t)} \sim p(\sigma_c^2|\bm{\tau}^{(t)}, \bm{\theta}^{(t)}, (\bm{\sigma}^2_{1:(c-1)})^{(t)}, (\bm{\sigma}^2_{(c+1):C})^{(t-1)}, \bm{L})$.
		\ENDFOR
		\STATE Sample $\kappa^{(t)}\sim p(\kappa|\bm{\tau}^{(t)})$.
		\FOR {$m=1$ to $M$}
		\STATE Use M-H to sample $\alpha_m^{(t)}\sim p(\alpha_m|\bm{\theta}^{(t)}, \bm{\alpha}_{1:(m-1)}^{(t)}, \bm{\alpha}_{(m+1):M}^{(t-1)})$.
		\ENDFOR
	\end{algorithmic}
\end{algorithm}

Each cycle of the algorithm generates a realization of a Markov chain, which gives approximate samples from the marginal posteriors after a successful burn-in process\cite{geman1984}.
We check that the acceptance rate of the Metropolis-Hastings sampler is roughly between 25$\%$ and 50$\%$ for adequate mixing of posterior samples \cite{gelman1996}.
The potential scale reduction $\hat{R}$ \cite{gelman1992} is also used to check convergence of the Markov chains.  We run the chains until $\hat{R}$ is less than 1.1 or 1.2, using  $\hat{R}$  of the logarithm of the posterior distribution as a benchmark.  A geometric decay of the autocorrelation as a function of the lag also suggests well mixing of our chains.  

We also conduct a simulation study to verify the M-w-G's ability to converge to the target distribution:

\begin{algorithm}
	\caption*{\textbf{Algorithm Example} Simulation to Verify Convergence of M-w-G}
	\begin{algorithmic}[1]
		\STATE Select the same four component aerosols as in the Beijing case studies  (Section \ref{sec:res}).
		\STATE $\kappa \leftarrow 100$ (or 500 for different runs).
		\STATE $\bm{\alpha} \leftarrow (0.8, 0.4, 0.2, 0.2)$ (or (2, 4, 0.1, 0.1) for different runs).   
		\STATE Sample $\bm{\tau}^{(0)} \sim$ \eqref{eq:tauprior}.
		\STATE Sample $\bm{\theta}^{(0)} \sim$ \eqref{eq:thetaprior}.
		\STATE Considering ($\bm{\tau}^{(0)}$, $\bm{\theta}^{(0)}$) as the true values, simulate radiances $\bm{L}^{sim}$ using the SMART lookup table and an additive Gaussian noise\footnotemark.
		\STATE Input $\bm{L}^{sim}$ into M-w-G retrieval algorithm to estimate AOD and mixing vectors.   
	\end{algorithmic}
\end{algorithm}
\footnotetext{The noise's standard deviation $\bm{\sigma}$ is set as 10\% of the averaged radiance, while the MISR operational algorithm estimates $\bm{\sigma}$ as 5\% of the same average.}

   The trace plots of the MCMC samples of AOD values (Figure \ref{fig:tautraces}\footnote{We attach in appendix two trace plots showing one example of each type, up to the first 1000 iterations.}) show good convergence after approximately 400 iterations, whether the initialization is close to the true value or not.  
	We observe similar convergence rates for mixing vectors. 
	Assigning different values to the hyperparameters, the correlation between the true AOD and the MCMC-retrieved AOD ranges between 0.78 and 0.90, and the coefficient of variation of the rooted-mean-square error ranges between 4.24\% and 9.26\%.  

Finally, we use the sampled posterior mean to estimate AOD values and mixing vectors. While the MCMC algorithm enables us to handle a hierarchy whose complexity precludes fitting by analytical methods, its computational intensity limits its operational use. Next, we propose a parallel MCMC  algorithm to reduce computational cost.

\subsection{A Parallel MCMC Algorithm}\label{sec:par}

Many MCMC sampling algorithms for spatial data suffer from high computational cost caused by the large dimensionality of data.  At 4.4 km resolution, our MCMC algorithm simulates samples for more than 16,000 variables\footnote{Excluding cloudy pixels can sometimes reduce the total dimensions to around 5,000 for one MISR Block.} for one MISR Block.  
	The large computational cost is exacerbated by the non-closed form of the posterior distributions.  
	It is possible, however, to develop a faster algorithm to sample from a distribution which approximates the target posterior of the original MCMC algorithm.

By this token, we devise a parallel MCMC algorithm to improve the computational efficiency: each MISR block is divided into 2 $\times$ 8 patches of equal size with at least four overlapping columns and rows for adjacent patches; the M-w-G sampler is applied to each patch independently to generate samples for ($\bm{\tau}$, $\bm{\theta}$). This independent sampling on different patches can therefore benefit from parallel computing.  

Information on AOD's spatial dependence structure is to be communicated across the entire MISR Block to estimate AOD's spatial smoothness level.
On that account, we let the patches periodically exchange spatial smoothness information across the entire MISR Block.  
Given $\kappa$'s conditional posterior,

\begin{equation*}
p(\kappa|\bm{\tau})\propto \kappa^{(P-1)/2-1}\exp\{-\frac{1}{2}\kappa\sum_{p\sim p^{\prime}}(\tau_{p}-\tau_{p^{\prime}})^2\},
\end{equation*}
and summary statistic, $T_{\kappa} = \sum_{p\sim p^{\prime}}(\tau_{p}-\tau_{p^{\prime}})^2$, it follows that $p(\kappa|\bm{\tau})=p(\kappa|T_{\kappa})$.
Hence $T_{\kappa}$ summarizes the information on calibration $\kappa$ for AOD's spatial smoothness across the entire MISR block. 
	Given that the hyperparameters control the spatial smoothness level of model parameters in all patches, the parallel MCMC algorithm provides an approximation to the posterior of AOD values $\bm{\tau}$, while the patch-samplings in parallel improve the computational efficiency. We now describe the parallel MCMC algorithm in detail:

\begin{algorithm}
	\caption*{\textbf{Parallel MCMC Algorithm}}
	Obtain a MISR Block of 32$\times$128 pixels at 4.4 km resolution and divide the Block into 2$\times$8 patches, each of 20$\times$20 pixels, with at least 4 overlapping columns/rows between adjacent patches. At step $t$, iterate the following process:
	\begin{algorithmic}[1]
		\STATE Use M-w-G algorithm to sample $\bm{\tau}\sim p(\bm{\tau}|\bm{\theta},\kappa,\bm{\sigma}^2,\bm{L})$, $\bm{\theta}\sim p(\bm{\theta}|\bm{\tau},\bm{\alpha},\bm{\sigma}^2,\bm{L})$, $\bm{\sigma}^2\sim p(\bm{\sigma^2}|T^{(t)}_{\bm{\sigma}})$, $\kappa\sim p(\kappa|T^{(t)}_{\kappa})$, $\bm{\alpha}\sim p(\bm{\alpha}|T^{(t)}_{\bm{\alpha}})$ within each patch in parallel for 50 iterations.
		\STATE Average the samples of the overlapping pixels between any two adjacent patches.
		\STATE Calculate summary statistics using current samples,
		\begin{center}
			$T_{\sigma_c}^{(t+1)} = \sum_{p=1}^P(L_{cp} - L_c^{RT}(\tau_p, \bm{\theta}_p))^2, c=1,\dots,C$,\\
			$T_{\kappa}^{(t+1)} = \sum_{p\sim p^{\prime}}(\tau_{p}-\tau_{p^{\prime}})^2$,\\
			 $T_{\alpha_m}^{(t+1)}= \sum_{p=1}^P\log\theta_{pm}, m=1,\dots, M$.
				\end{center}
		\end{algorithmic}
	\end{algorithm}

\noindent The above process can be automated using the Perl programming language. For a MISR block at 4.4~km resolution, the computational time of the parallel MCMC algorithm is less than one-fifth of that of the global MCMC sampling algorithm, accounting for overhead time of communication among different patches.

This parallel MCMC sampling scheme can be generalized to improve the computational efficiency of MCMC sampling based on spatial data of a large scale.  
	By conditioning on a summary statistic which preserves the global spatial dependence level, we can partition the original sampling problem into many sub-samplings and distribute them to different processing units concurrently.  Samples generated from each processing unit can be periodically collected to renew the summary statistic, which is then returned to each processing unit to update the sub-samplings.
	Though this scheme samples from an approximation to the target distribution, it can largely speed up the computation.

The global and the parallel MCMC  algorithms  produce reasonably consistent results. The outputs generally agree, except for a small group of pixels that mostly lie on the patch edges.
The spatial smoothness is interrupted between patches; the benefits of a stabilizing factor from neighboring pixels are lost.  
This confirms that maintaining an appropriate spatial structure  is  important, and that our parallel MCMC algorithm's outputs are only an approximation to the target distribution.  
Increasing the number of iterations and communications of summary statistics, and smoothing the patch edges, reduce the disagreement.
The next section evaluates the performance of our retrievals using case studies.  For non-operational model validations, we apply the global MCMC algorithm to avoid inconsistency in number of iterations for different retrievals.

\section{Validation and Results: Case Studies on Aerosol Retrievals for the Greater Beijing Area, China}\label{sec:res}
 
In this section, we compare our retrievals to MISR outputs for the greater Beijing area (latitude: 38.95N$\sim$40.15N; longitude: 115.57E$\sim$119.50E) and discuss their differences.  
We validate our results using AERONET measurements and Google Earth satellite images. 
Through case studies, we demonstrate the importance of fine-resolution retrievals and a greater variety of compositions to improve retrieval accuracy and coverage.

\subsection{Comparison with MISR Retrievals}\label{sec:wmisr}

\begin{figure}[ht]
\centering
\resizebox{0.45\columnwidth}{!}{\includegraphics{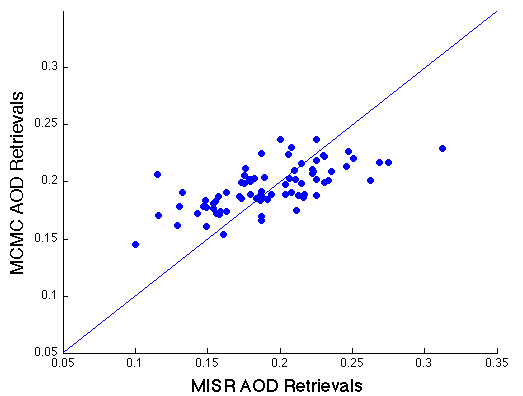}}
\resizebox{0.45\columnwidth}{!}{\includegraphics{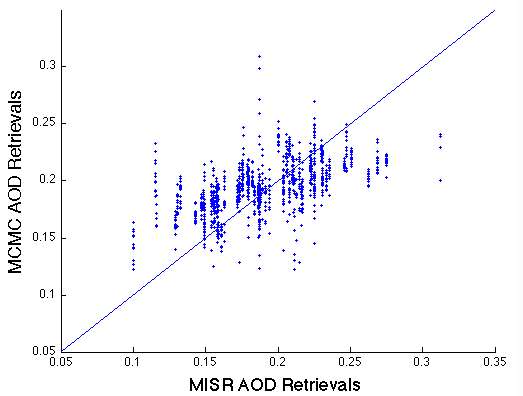}}
\caption{\label{fig: bench}Scatterplots of MISR against MCMC retrievals  at an aggregated 17.6 km resolution (left, r.m.s. = 0.0295) and a 4.4 km resolution (right, r.m.s. = 0.0309).}
\end{figure}

Figure \ref{fig:compare} displays the MISR AOD retrievals at 17.6 km resolution in panel (a) and our Bayesian AOD retrievals at 4.4 km resolution in panel (b). 
Shared information in MISR and our retrievals is observed, such as the coastline on the right, the overall AOD level, and its spatial patterns.  This consistency is confirmed by the scatterplots of MISR outputs and Bayesian AOD retrievals aggregated to 17.6 km resolution (Figure \ref{fig: bench}, left panel).
The black pixels in Figure \ref{fig:compare} represent missing retrievals, mostly due to two common reasons.  Firstly,
aerosol retrievals are not attempted when clouds are detected.  
{\footnotesize 
\begin{figure}[htp]
\centering
\resizebox{0.7\columnwidth}{!}{\includegraphics{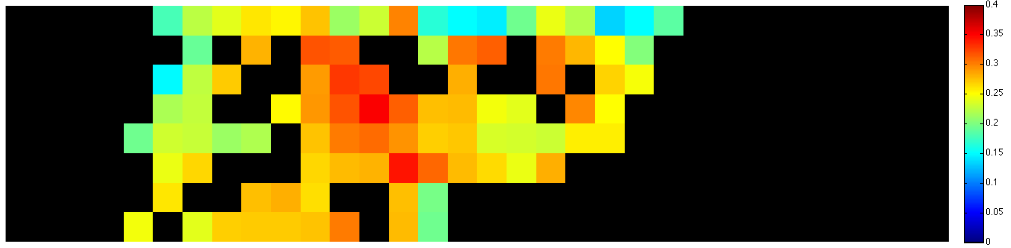}}\\
{\footnotesize (a) MISR AOD retrievals at 17.6 km resolution.}
\resizebox{0.7\columnwidth}{!}{\includegraphics{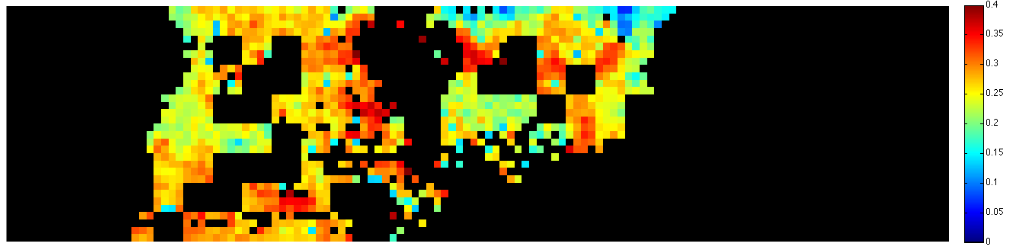}}\\
{\footnotesize (b) Bayesian AOD retrievals using MCMC at 4.4 km resolution.}
\caption{\label{fig:compare} AOD estimates from MISR and our Bayesian retrievals.}
\end{figure}}	MISR averages the 1.1 km observations into a pixel at 17.6 km resolution and ignores clouds, when the cloudless areas are more than $\frac{1}{16}$ of the pixel.  
	Clouds negligible at 17.6 km resolution, however, might be significant at 4.4 km resolution; we tend to have more missing retrievals in some areas, intrinsically determined by the observations. 
	Secondly, when none of the 74 MISR-designated compositions satisfy MISR's weighted least squares criterion, MISR operational algorithm marks the retrieval as missing.  Our Bayesian retrievals, allowing for a richer variety of compositions, eliminate such unnecessarily missing retrievals (Section \ref{sec:comp}).

On the other hand, Figure \ref{fig:compare} also demonstrates increased diversity in our Bayesian-retrieved AOD across the MISR Block, as the retrieval resolution improves. This is expected, since a finer resolution leads to more information observed and piped into the model.  
The reliability of such diversity needs to be further validated by other independent sources such as ground-based measurements, as discussed in the next section.

\subsection{\label{sec:ground}Model Validation for Bayesian Retrievals by Ground-based Measurements and Google Earth}

Ground-based measurements are collected at AERONET Beijing and AERONET Xianghe stations, as well as via a hand-held MICROTOPS II Sunphotometer at several locations in urban Beijing area.  The fixed locations of the AERONET stations and the limited travel range of the Sunphotometer's human operator make it impossible to validate retrievals of all pixels on the MISR Block under study.  
	Instead, we focus on the pixels that contain the AERONET stations or our Sunphotometer-visited locations.  To match the AOD values at the same wavelength, we first convert AERONET measurements to those at 550 nm using AERONET estimates of Angstr\"{o}m exponent.  We then average the measurements within a one-hour  window when Terra carrying MISR passes over the AERONET stations, for Jiang, \textit{et al.} \cite{jiang2007} show that a narrower time window better captures the correlation between AERONET measurements and MISR retrievals.
The area's frequent cloudy weather and its latitude\footnote{The Beijing city is visited by the Terra satellite every five to nine days.} contribute to the scarcity of  the remote-sensed versus ground-based data pairs for  validation.  

As a result of this scarcity of ground-based validation, we also carry out qualitative validation using satellite images from Google Earth and discuss the findings in Section \ref{sec:googleear}.

\subsubsection{\textbf{Retrieval Validation at AERONET Beijing Station}}\label{sec:validbj}

\begin{figure}[htp]
\centering
\resizebox{0.75\columnwidth}{!}{\includegraphics{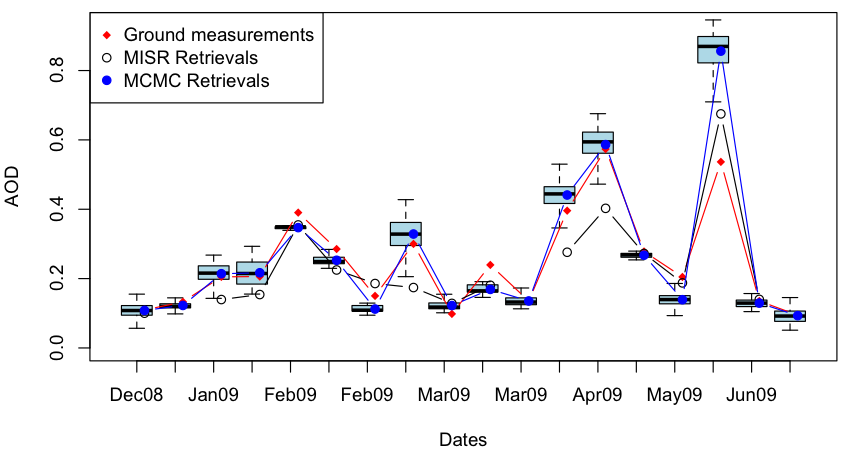}}
\includegraphics[scale=0.41]{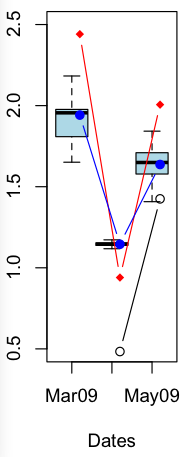}
\caption{\label{fig:valid_bj}Validation of our AOD retrievals by measurements at AERONET Beijing Station.}
\end{figure}

Figure \ref{fig:valid_bj} shows a boxplot of our Bayesian AOD retrievals for the pixel which contains the AERONET Beijing Station\footnote{Latitude: 39.97689$^{\circ}$ North; longitude: 116.38137$^{\circ}$ East.}, 
with estimated uncertainties indicated by the box edges for inter-quartile ranges of posteriors and the whiskers drawn to the 5th and 95th percentiles.  
The three retrievals on March 15, April 30, and May 16, 2009 are plotted separately in the right panel to keep an appropriate scale for the left panel.

As long as a pixel is cloudless, our MCMC algorithms provide an AOD retrieval.  However, the MISR operational retrieval algorithm shows missing values for 24\% of the 21 cases in Figure \ref{fig:valid_bj}. 
This results from the increasingly heterogeneous aerosol conditions in Beijing and the limited choices of aerosol compositions in MISR retrievals.
In the coarse-resolution retrievals, 
high AOD values are averaged down by its neighbors and low AOD values averaged up, resulting in a loss of useful information.  Our Bayesian retrievals show improvement in accuracy; detailed information on aerosols are  revealed by the fine-resolution retrievals.  The three high AOD values in the right panel of Figure \ref{fig:valid_bj}  indicate Beijing's extreme air conditions, corresponding to 86$\%$, 71$\%$, and 81$\%$ reduced radiation by aerosols.  For example, records of news from Xinhua Headlines show that on March 15, 2009, the city was trapped in a sandstorm originated in Inner Mongolia.

We would like to discuss one particular case when the Bayesian retrieval (0.8560) is much worse than MISR output (0.6750), compared to the AERONET measurement (0.5455): the third to last case in Figure \ref{fig:valid_bj} (left panel), May 25, 2009.  AERONET reports no measurement when  Terra carrying MISR passed above Beijing. Instead, we use the measurement of 0.5455, which is the closest in time but three hours earlier.  This record reached the lowest of that day, with others between 0.6521 and 1.5366.  It suggests that the particular AERONET record we used might not be ideal to validate the remote-sensed retrieval, but our best option.

\subsubsection{\textbf{Retrieval Validation at AERONET Xianghe Station}}\label{sec:validxh}

\begin{figure}[htp]
\centering
\resizebox{0.76\columnwidth}{!}{\includegraphics{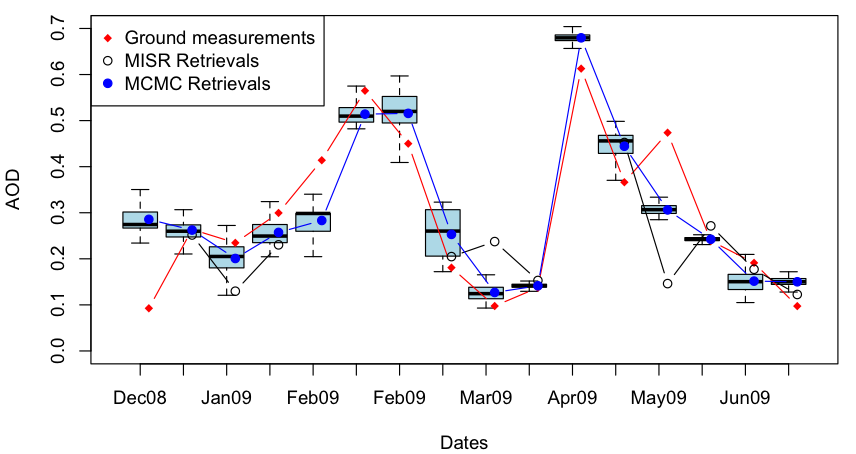}}
\includegraphics[scale=0.41]{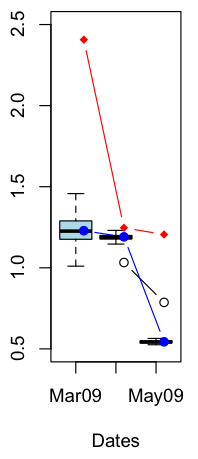}
\caption{\label{fig:valid_xh}Validation of our AOD retrieval results by AERONET measurements in Xianghe.}
\end{figure}

Figure \ref{fig:valid_xh} compares the remote-sensed retrievals to AERONET measurements at the pixel that contains the AERONET Xianghe station\footnote{Latitude: 39.75360$^{\circ}$ North; longitude: 116.96150$^{\circ}$ East.}.  From December to February, AERONET measurements are mostly higher than remote-sensed retrievals, but no distinctive pattern afterwards. 

AERONET Xianghe station has the Jingshen Expressway to its north, which is a major path connecting two hub cities: Beijing and Shenyang\footnote{The capital and largest city of Liaoning Province in Northeast China.}. 
The northwest wind in winter carries  car exhaust to the AERONET Xianghe station, possibly leading to high AOD measurements.  Yet for remote-sensing retrievals, the green fields in a larger neighborhood balance this factor, which possibly results in a washed-out signal. 
However, the fine-resolution retrievals seem to suffer less from the balancing factors and display a better accuracy.

We would like to discuss one of the cases where our AOD retrieval is much higher than AERONET measurement: the first data point in Figure \ref{fig:valid_xh}, December 25, 2008.  MISR  produced no output for this day.  To the east of Xianghe station  in Hebei Province lie several major malls for furniture exhibition and manufacture.  On December 25, 2008,  the furniture companies started renovating their exhibition halls. 
The construction could have caused  localized aerosol loadings not observed by the Xianghe AERONET site 2~km away upwind within one day, but detected by the MISR instrument and captured by our retrievals.  

\subsubsection{\textbf{Qualitative Validation using Google Earth in Absence of Ground Measurements}}\label{sec:googleear}

	 We observe other disagreements in our Bayesian-retrieved AOD values and those of MISR, in addition to those at the two pixels that contain the two AERONET stations in the MISR Block. Since they are retrievals at different spatial scales, they could as well be different, that is, they could both be but both valid.
	An indirect way to validate our retrieved AOD values is to see whether they reasonably reflect the region's geographical and anthropogenic conditions, such as existence of heavy industries and transportation patterns. 
	These conditions can be easily assessed using the satellite images from Google Earth, making them indirect validation for our retrievals as a reasonable and detailed 
profiling of  AOD spatial distribution.
	Here we focus on pixels with Bayesian AOD retrievals largely disagreeing with those of their adjacent pixels or pixels with locally highly variable Bayesian AOD retrievals.
	Since our retrieval pixels are only $\frac{1}{16}$ of the size of a MISR retrieval pixel, these AOD locations cannot be identified in the corresponding MISR retrievals.

\begin{figure}[htp]
\centering
\includegraphics[scale=0.2]{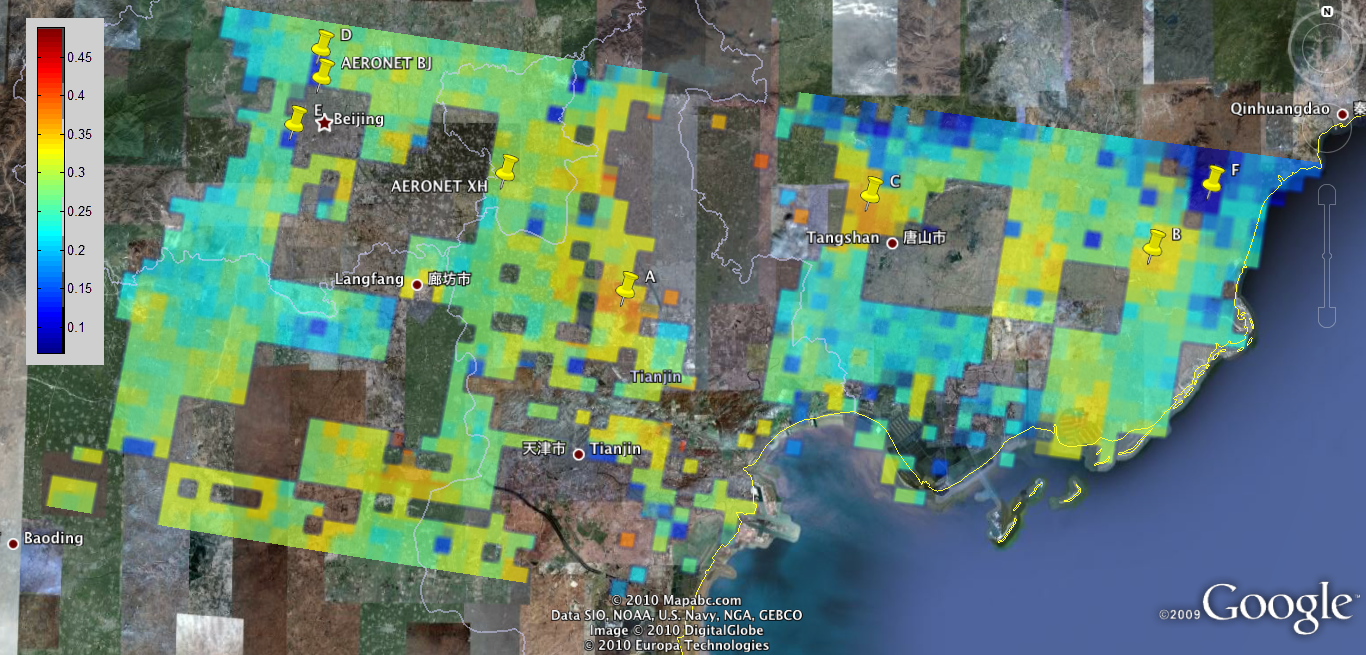}
\caption{Retrieval results projected on Google Earth.}\label{fig:google}
\end{figure}

	In particular, we project our Bayesian AOD values onto Google Earth (Figure \ref{fig:google}) and examine the pixels with locally highly variable AOD values.
We thus identify a hub of the Jingshen and Jingtang Highways (pin A in Figure \ref{fig:google}) and construction sites producing pollution (pin C), supporting the high AOD values indicated by only our Bayesian retrievals.   The Olympic Park (pin D) and Beidaihe (pin F), a famous beach resort, also confirm the reasonability of the low AOD values captured by only the Bayesian retrievals at a finer resolution.

\subsection{\label{sec:comp}Case Study for Including a Richer Variety of Aerosol Compositions}

This section emphasizes the necessity to expand MISR's 74 aerosol compositions.  This expansion improves retrieval coverage and detects more features of aerosol behaviors, such as seasonality of component aerosols.

\begin{figure}[htp]
\centering
\includegraphics[scale=0.4]{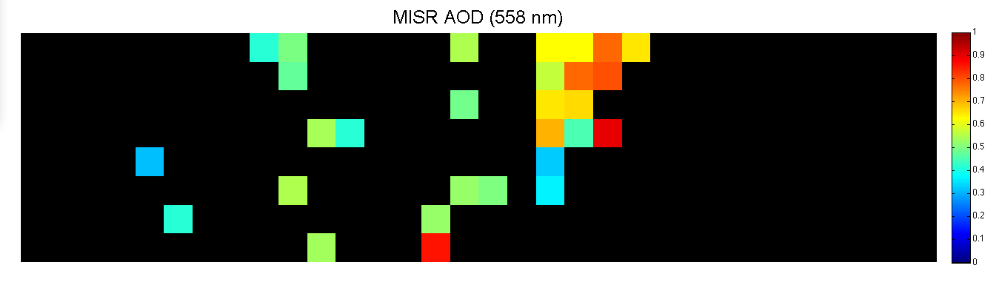}\\
{\scriptsize MCMC AOD (558nm)}
\includegraphics[scale=0.4]{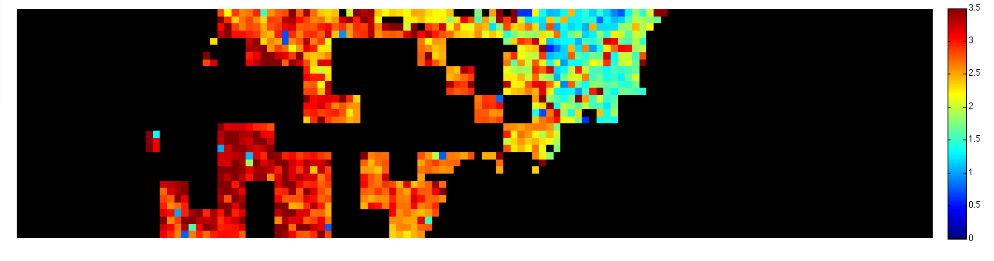}
\caption{Case study of AOD retrievals on March 15, 2009.}\label{fig:map}
\end{figure}

For an example of the improvement on retrieval coverage, we examine March 15, 2009.  MISR failed to retrieve AOD for the majority of the Block (Figure \ref{fig:map}, upper panel).  Our Bayesian retrievals provide better coverage and the retrievals give distinctly high AOD values with a clear path of aerosols migrating from west to east and into the ocean.  This  unusual discrepancy leads us to run through the weather records: on that day, the area suffered from a sandstorm originated in Inner Mongolia, which later passed into eastern China.
For areas like Beijing, which experience occasional sandstorms, the limited compositions containing grains(dust) among MISR's 74 choices could easily result in a low coverage of MISR retrievals.  Similar situations might exist for other locations with usual aerosol conditions.
The retrieved mixing vectors also contains information on the regional aerosol composition and can be used to identify pollution type and source.  For example, results show that component No.6 with sulfate tends to dominate the composition in winter due to coal burning for heating, with No.19, grains (dust) dominating in spring due to sandstorms. 

\begin{figure}[htp]
\centering
\includegraphics[scale=0.55]{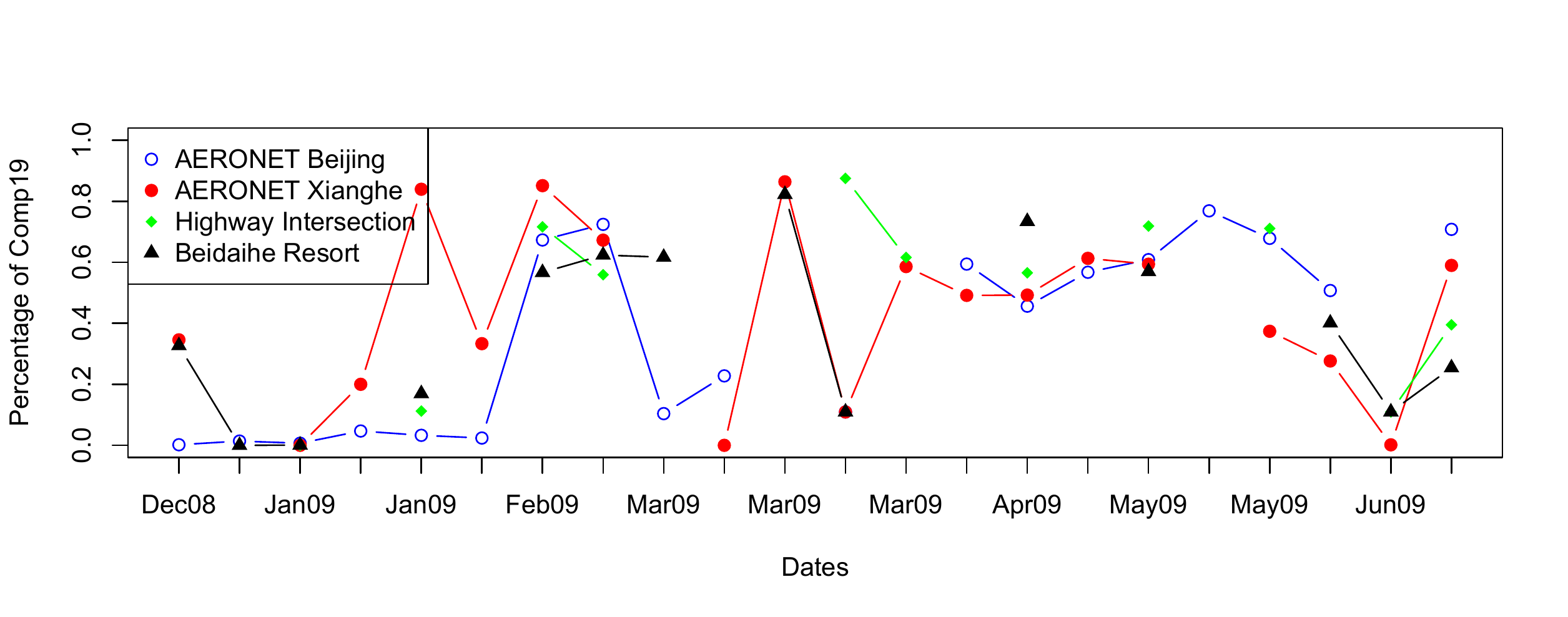}
\caption{Mixing percentages of component No.19 from winter to spring.}\label{fig:compsea}
\end{figure}

Figure \ref{fig:compsea} shows the mixing percentages of component No.19 over December 2008 to June 2009, at four different locations: the AERONET Beijing station, the AERONET Xianghe station, location (A) and (F) marked in Figure \ref{fig:google}.  
For AERONET Beijing station, the percentage of grains(dust) only rose in the spring, due to the sandstorms, while the constructions around AERONET Xianghe might have raised the percentage earlier in the year.  Location (A), where major highways intersect, showed a high amount of  dust in its aerosol compositions through the warm seasons when traffic typically increases.  The mixing percentage of No.19 at Location (F), the Beidaihe Resort, moved relatively in  consistence with AERONET Beijing station.  We hope to  explore this  trend in future research.
In general, by correctly identifying the major pollutants for each season, we can better understand the transitions of aerosols and, therefore, take efforts to improve air quality in a more specific and to the point manner.  For accuracy and coverage, it is necessary to expand the MISR-designated 74 aerosol compositions to a richer variety.

\section{Discussion}\label{sec:dis}

Aerosols serve as an important factor in air quality and public health.  
A  profile of AOD's spatial distribution can eventually expand the potential of remote-sensed observations in facilitating urban air quality monitoring and public health studies \cite{tinkle2007}\cite{tatem2004}.   
The heterogeneity of urban aerosols due to anthropogenic activities calls for a profile of aerosols at a fine resolution and a larger variety of aerosol compositions.

In this paper, we have presented a hierarchical Bayesian  model to retrieve AOD values and mixing vectors relative to a collection of four component aerosols at an improved resolution of 4.4 km using MISR observations. 
The model incorporates a spatial dependence structure to gain strength from AOD's spatial smoothness; it also allows for a richer variety of aerosol mixing vectors to better capture the growing heterogeneity of urban aerosols and the increasingly severe weather conditions, such as sand storms.  
A more detailed AOD spatial profile is provided and further validated by AERONET and Google Earth; an improved accuracy and a better retrieval coverage is obtained due to the improved resolution and flexible choices of aerosol compositions.  This improvement is particularly important during high-AOD events, which often indicate severe air pollution.  We further develop a parallel MCMC algorithm to improve the computational efficiency, which can be generalized to speed up other MCMC sampling algorithms based on spatial data.

From the case studies, we become more aware of the complexity in aerosol conditions and thus hope to use our results to study the aerosols' impact on public health in urban areas at the enhanced resolution.  
We also hope to explore the possibility of improving the retrieval accuracy by incorporating more prior knowledge in the model, such as wind measurements and dependence among the four spectral bands.

\section{Acknowledgments}

	The authors gratefully acknowledge support from the National Science Foundation Grants DMS-0907632, DMS-1107000, SES-0835531 (CDI)
and CCF-0939370, and ARO grant W911NF-11-1-0114, the National Science Foundation of China (60325101,
60872078), Key Laboratory of Machine Perception (Ministry of Education) of Peking University, and Microsoft Research of Asia.
	The authors would like to thank Dr. Susan Paradise, Dr. Amy Braverman, and the MISR team for their great support, Derek Bean for editing advice.  We also thank the AERONET PIs, Hong-Bin Chen, Philippe Goloub, Pucai Wang, Zhanqing Li, Brent Holben, and Xiangao Xia for establishing and maintaining the Beijing and Xianghe AERONET stations, especially Pei Wang from Peking University for collecting AOD data in Beijing.  We thank Graham Shapiro for implementing the projection of AOD retrievals to Google Earth.  We thank Dr. Chengcai Li from Peking University and Yang Liu from Emory University for discussions.  
	Last but not least, we would like to thank Hal Stern, Editor of JASA's Applications and Case Studies, the anonymous Associate Editor and Referee for their thoughtful comments and suggestions that helped us improve our work and manuscript.

\section{Appendix: Example trace plots of MCMC samples for AOD in simulation study (Section 3.2)}
\begin{figure}[htp]
  \centering		    \resizebox{0.85\columnwidth}{!}{\includegraphics{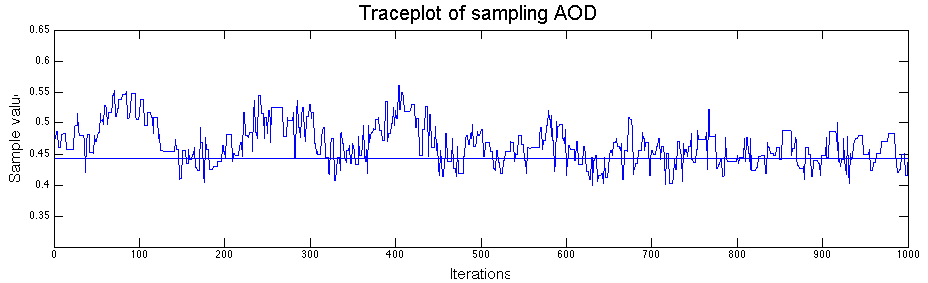}}	\resizebox{0.86\columnwidth}{!}{\includegraphics{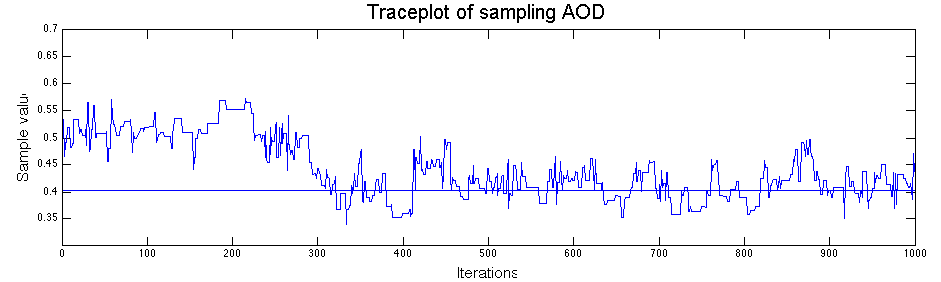}}
					\caption{Example of sampling trace plots of AOD retrievals.}\label{fig:tautraces}
\end{figure}

\newpage

\bibliographystyle{ECA_jasa}
\bibliography{yqrefs}


\end{document}